\documentclass[pra,twocolumn,tightenlines,nofootinbib]{revtex4}
\usepackage{graphicx}
\usepackage{multirow}
\usepackage{amsmath,amsfonts,amssymb}
\usepackage{bm}
\usepackage{dcolumn}
\usepackage{graphicx}

\newcommand{\eref}[1]{Eq.~(\ref{#1})}

\begin{document}

\title{Multipolar polarizabilities and hyperpolarizabilities
in the Sr optical lattice clock}

\author{S.~G.~Porsev$^{1,2}$}
\author{M.~S.~Safronova$^{1,3}$}
\author{U.~I.~Safronova$^{4}$}
\author{M. G. Kozlov$^{2,5}$}

\affiliation{
$^1$Department of Physics and Astronomy, University of Delaware, Newark, Delaware 19716, USA\\
$^2$Petersburg Nuclear Physics Institute, Gatchina, Leningrad District, 188300, Russia\\
$^3$Joint Quantum Institute, National Institute of Standards and Technology and the University of Maryland,
Gaithersburg, Maryland 20742, USA \\
$^4$Physics Department, University of Nevada, Reno, Nevada 89557 \\
$^5$St.~Petersburg Electrotechnical University ``LETI'', Prof. Popov Str. 5, 197376 St.~Petersburg}
\date{\today}

\begin{abstract}
We address the problem of the lattice Stark
shifts in the Sr clock caused by the multipolar $M1$ and $E2$ atom-field interactions and  by the term nonlinear in lattice
intensity and determined by the hyperpolarizability. We have developed an approach to calculate
hyperpolarizabilities for atoms and ions based on a solution of the inhomogeneous equation which allows to
effectively and accurately carry out complete summations over intermediate states. We
applied our method to the calculation of the hyperpolarizabilities for the  clock states in Sr. We also carried out an accurate calculation of the multipolar polarizabilities for these states at the magic frequency. Understanding these
Stark shifts in optical lattice clocks is crucial for further improvement of the clock accuracy.
\end{abstract}
\pacs{ }

\maketitle
\section{Introduction}
Extraordinary advancements in the optical atomic clock accuracy and stability have been demonstrated in the past few years \cite{LudBoyYe15,UshTakDas15,NicCamHut15,HunSanLip16,SchBroMcG17,CamHutMar17}.
The systematic  uncertainty of the Sr optical lattice clock was reduced to $2.1\times 10^{-18}$ in fractional frequency units~\cite{NicCamHut15}. Similar systematic uncertainty of $3.2\times 10^{-18}$ was reported in a single trapped ion
atomic clock based on the electric-octupole transition in Yb$^+$~\cite{HunSanLip16}.
A fractional frequency instability of $6\times 10^{-17}/\sqrt{\tau}$ for an averaging time $\tau$ in seconds was demonstrated in Ref.~\cite{SchBroMcG17} using a zero-dead-time optical clock based on the interleaved interrogation of two cold-atom ensembles.
A Fermi-degenerate three-dimensional optical lattice clock was demonstrated in \cite{CamHutMar17}, with a synchronous
clock comparison between two regions of the 3D lattice yielding a measurement precision of
$5\times 10^{-19}$ in 1 hour of averaging time. Improved precision of the atomic clocks
enables many applications, including relativistic geodesy \cite{Flu16}, very long baseline interferometry \cite{NorCle11},
search for the variation of the fundamental constants \cite{SafBudDem17} and dark matter \cite{Arv15,RobBleDai17},
tests of the Lorentz invariance \cite{PihGueBai17}, redefinition of the second \cite{BreMilPiz17}, and others. Further improvement of clock precision is needed for these applications and implementation of new ideas, such as the use of atomic clocks for the gravitational wave detection \cite{KolPikLan16}.

When an atom is placed in a laser field, the atomic energy levels shift due to the ac Stark effect. In the main approximation, these shifts are determined by the frequency-dependent electric dipole polarizabilities of the atomic states \cite{MitSafCla10}.
To cancel the ac Stark shift of the clock transition, the trap laser of the optical lattice clock
is operated at the ``magic wavelength'' \cite{KatIdoKuw99,YeVerKim99}, at which the electric dipole polarizabilities of the clock states are the same, resulting in zero ac Stark shift of a clock transition  to some degree of precision.

With the  clock systematic uncertainties reaching $10^{-18}$, such cancellation is no longer possible as other contributions to the Stark shift effects become significant \cite{OvsPalTai13,KatOvsMar15,BroPhiBel17}: the magnetic-dipole ($M1$) and electric-quadrupole ($E2$) interactions of the atom with the lattice field and
the term nonlinear in the trap laser intensity and determined by the hyperpolarizability.

Both measurements and calculations of these effects are very difficult \cite{BroPhiBel17}, with contradictory results obtained for Sr
\cite{KatOvsMar15,OvsPalTai13,NicCamHut15,WesLodLor11,LeTLorLeC13} for both theory and experiment. Theoretical evaluation of the $M1$
and $E2$ polarizabilities in Sr disagree even in the sign of the effect \cite{KatOvsMar15,OvsPalTai13}.  The experimental determination of the Sr clock differential hyperpolarizability $\Delta \beta = \beta(^3\!P_0^o) - \beta(^1\!S_0)$
at the magic frequency by JILA ~\cite{NicCamHut15} is consistent with zero, $\Delta \beta=-1.3(1.3)\times 10^{7}$ a.u., but the
SYRTE~\cite{WesLodLor11,LeTLorLeC13} measurement $-2.01(45)\times 10^{7}$ a.u. is not.
Recent experimental assessment of the nonlinear hyperpolarizability ac Stark shifts in Yb found a difficulty associated with the finite temperature effects \cite{BroPhiBel17}.
In summary, reliable theoretical values of the ac  multipolar polarizabilities and hyperpolarizability are required to control lattice shifts at the $10^{-19}$ level.

So far there are no reliable calculations of the clock states hyperpolarizabilities for divalent atoms.
An accurate calculation of hyperpolarizabilities is very difficult because
certain terms are expressed by complicated formulas, involving three summations over intermediate states, which cannot be properly evaluated by summing over low-lying contributions.
An exception is the case of Yb, where the magic wavelength happens to be accidentally close to relevant atomic transition energies causing strong resonant effects, and making a very few hyperpolarizability terms dominant. For Sr, however, full expression must be evaluated, and we focus on this example.

In this work we develop a method that allows us to solve this problem and accurately calculate the hyperpolarizabilities
for states with the total angular momentum $J=0$ which includes Sr, Yb, Hg, Mg, and Cd clock cases. A generalization of the method to higher total angular momenta is straightforward but technically more complicated.
We applied our method to calculate the $^1\!S_0$ and $^3\!P_0^o$ clock states
hyperpolarizabilities at the magic frequency in Sr and their uncertainties. Our results definitively determine that the Sr
hyperpolarizability is not zero and has to be taken into account when reducing lattice shifts at the next level of precision.
We also calculated Sr clock magnetic-dipole and electric-quadrupole polarizabilities, resolving previous major discrepancies and providing the final differential multipolar polarizability with only 13\% uncertainty.
\paragraph{General formalism.}
We assume that an atom in a state $|0\rangle \equiv |\gamma_0, J_0=0 \rangle$ (where $\gamma_0$ encapsulates all other quantum numbers), is placed in a linear polarized field of the lattice standing wave, given by 
\begin{equation}
{\mathcal E}_z = 2 {(\mathcal E_0)}_z \, {\rm cos}(kx)\, {\rm cos}(\omega t).
\end{equation}
Here $k=\omega/c$, $\omega$ is the lattice laser wave frequency, $c$ is the speed of light, and the factor 2 accounts for the superposition of forward and backward traveling waves that doubles the lattice standing-wave amplitude.
A spatial part of the atom-lattice interaction is determined as~\cite{OvsPalTai13,KatOvsMar15}
\begin{equation}
̂ V(x)= V_{E1}\, {\rm cos}(kx) + (V_{M1}+V_{E2})\, {\rm sin}(kx)
\end{equation}
where $V_{E1}$, $V_{M1}$, and $V_{E2}$ correspond to operators of $E1$, $M1$, and $E2$ interactions
and $x$ determines the position of the atom starting from the standing-wave antinode.

The optical lattice potential for the atom at $|kx| \ll 1$ can be approximated as~\cite{OvsPalTai13,KatOvsMar15}
\begin{eqnarray}
U(\omega) &\approx& -[\alpha^{E1}(\omega) -\{\alpha^{M1}(\omega) + \alpha^{E2}(\omega)\}k^2 x^2]\,\mathcal{E}_0^2 \notag \\
&& -\beta(\omega)\, \mathcal{E}_0^4 .
\label{DeltaE}
\end{eqnarray}

The ac $2^K$-pole polarizability of the $|0\rangle$ state is expressed
(we use atomic units $\hbar=m=|e|=1$) as~\cite{PorDerFor04}
\begin{eqnarray}
\alpha^{\lambda K}(\omega) &=& \frac{K+1}{K} \frac{2K+1}{[(2K+1)!!]^2}
(\alpha \,\omega)^{2K-2}  \notag \\
&\times& \sum_n \frac{(E_n-E_0) | \langle n||T_{\lambda K}||0 \rangle |^2}{(E_n-E_0)^2-\omega^2} .
\label{Qlk}
\end{eqnarray}
Here $\lambda$ distinguishes between electric, $\lambda = E$, and magnetic, $%
\lambda =M$, multipoles and $\langle n||T_{\lambda k}||0 \rangle$ are
 the reduced matrix elements of the multipole operators,
$T_{E1} \equiv D$, $T_{M1} \equiv \mu$, and $T_{E2} \equiv Q$.

The hyperpolarizability $\beta(\omega)$ of the $|0\rangle$ state is determined as
\begin{equation}
\beta (\omega) = \frac{1}{9}\,Y_{101}(\omega) + \frac{2}{45} Y_{121}(\omega) .
\label{beta}
\end{equation}
The quantities $Y_{101}\left( \omega \right) $ and $Y_{121}\left( \omega \right) $ are
given by
\begin{eqnarray*}
\label{Y101}
Y_{101}(\omega) &\equiv& \sum_q \mathcal{R}_{101} (q\omega,2q\omega,q\omega) \notag \\
&+& \sum_{q,q'} \left[ \mathcal{R}'_{101}(q\omega,0,q'\omega)
- \mathcal{R}_1(q'\omega) \mathcal{R}_1(q\omega,q\omega) \right], \notag \\
Y_{121}(\omega)\! &\equiv& \!
\sum_q \left[ \mathcal{R}_{121} (q\omega,2q\omega,q\omega)
 + \sum_{q'} \mathcal{R}_{121}(q\omega,0,q'\omega) \right],
\end{eqnarray*}%
where $q,q'= \pm 1$ and
\begin{widetext}
\begin{eqnarray}
\label{R_J1}
\mathcal{R}_{1 J_n 1}(\omega_1,\omega_2,\omega_3) &\equiv&
\sum_{\gamma_m,\gamma_n,\gamma_k}
\frac{\langle 0 ||D|| \gamma_m, 1  \rangle \langle \gamma_m, 1 ||D|| \gamma_n, J_n \rangle
\langle \gamma_n, J_n ||D|| \gamma_k, 1 \rangle \langle \gamma_k, 1 ||D|| 0 \rangle}
{(E_m - E_0 -\omega_1) (E_n - E_0 -\omega_2) (E_k - E_0 -\omega_3)}, \\
\label{R_J3}
\mathcal{R}_1(\omega) &\equiv& \sum_{\gamma_m} \frac{|\langle 0 ||D|| \gamma_m, 1 \rangle|^2}{E_m-E_0-\omega},
\qquad
\mathcal{R}_1(\omega,\omega) \equiv \sum_{\gamma_k}\frac{|\langle 0 ||D|| \gamma_k, 1 \rangle|^2}{(E_k-E_0-\omega)^2} .
\end{eqnarray}
\end{widetext}
The notation $\mathcal{R}^{\prime}_{101}$ (i.e., the prime over $\mathcal{R}$) means that
the term $|\gamma_n, 0 \rangle = |\gamma_0, 0 \rangle$ must be excluded from the summation over $\gamma_n$ in Eq.~(\ref{R_J1}).

The term $\mathcal{R}_{1J_{n}1}(\omega_1,\omega_2,\omega_2)$ contains three summations over
intermediate states which so far precluded accurate calculation of the hyperpolarizabilities for lattice clocks.
A large number of the intermediate states should be included in the summation unless some accidental resonances arise making such approach impractical.

In this work we develop another method of calculating hyperpolarizabilities which is based on solutions of the inhomogeneous equations.
We consider
\begin{equation}
\alpha_{q\omega}(\gamma_0,0) =  \frac{2}{3} \sum_{\gamma_m}
\frac{\langle 0 ||D|| \gamma_m,1 \rangle \langle \gamma_m,1 ||D|| 0 \rangle}{E_0 + q\omega -E_m},
\label{alphaE0}
\end{equation}%
which can be treated as the scalar static electric dipole polarizability of
the $|0 \rangle$ state, calculated at the shifted energy $(E_0 + q\omega)$.
For brevity we omit the superscript $E1$, i.e., $\alpha_{q\omega} \equiv \alpha^{E1}_{q\omega}$.
We use the Sternheimer~\cite{Ste50} or
Dalgarno-Lewis~\cite{DalLew55} method and solve the inhomogeneous equation
\begin{eqnarray}
(E_0 -H_{\rm eff} + q\omega)\, |\delta \phi \rangle = D_z\, |0 \rangle,
\label{a5}
\end{eqnarray}
where $H_{\rm eff}$ is the effective Hamiltonian determined below (see~\eref{Heff}).
Solving \eref{a5} and introducing $|\delta \phi_q \rangle$ as
\begin{equation*}
|\delta \phi_q \rangle  \equiv \sqrt{3}\, |\delta \phi \rangle =
\sum_{\gamma_m} \frac{|\gamma_{m},1\rangle \langle \gamma_{m},1||D|| 0 \rangle}{E_0-E_m + q \omega} ,
\end{equation*}%
we find $\alpha_{q \omega}(\gamma_0,0)$ as the expectation value
\begin{eqnarray}
\alpha_{q \omega}(\gamma_0,0) = \frac{2}{3} \langle 0 ||D|| \delta \phi_q \rangle .
\label{a6}
\end{eqnarray}
Substituting $|\delta \phi_q \rangle$ into Eq.~(\ref{R_J1}), we obtain
\begin{eqnarray}
\mathcal{R}_{1 J_n 1}(q\omega, 2q\omega,q\omega)  
&=& \sum_{\gamma_n} \frac{|\langle \delta \phi_q ||D|| \gamma_n,J_n \rangle|^2}{E_n - E_0 - 2q\omega} .
\label{R_1Jn1}
\end{eqnarray}%
Thus, using this approach, we included all discrete
and continuum intermediate states in the sums over $\gamma_m$ and $\gamma_k$ in Eq.~(\ref{R_J1}).
To sum over all intermediate states at fixed values $J_n=0,2$ in Eq.~(\ref{R_1Jn1}) we apply again the method
of solution of inhomogeneous equation, as described above.

The calculation of the terms $\mathcal{R}'_{101} (q\omega,0,q\omega)$ is carried out in a similar way.
For example, for $\mathcal{R}'_{101} (\omega, 0, \omega)$ we arrive at
\begin{equation}
\mathcal{R}'_{101}(\omega,0,\omega) =
{\sum_{\gamma_n}}' \frac{|\langle \delta \phi_{+1} \,||D||\, \gamma_n,0\rangle|^2}{E_n-E_0}.
\label{R101}
\end{equation}
However, there is a complication in comparison to the previous case: we need to exclude the
term $|\gamma_n,0 \rangle =|\gamma_0,0 \rangle $ from the
summation over $\gamma_n$. We developed the following method to eliminate this term from the solution of the inhomogeneous equation.  We add a small imaginary term $i\omega^{\prime}$
to the denominator of~\eref{R101}, so that
\begin{equation*}
\mathcal{R}'_{101}(\omega,0,\omega) =
\mathrm{Re} \left\{\mathcal{R}'_{101}(\omega,i\omega' \rightarrow 0,\omega) \right\} ,
\end{equation*}
and find the real part of $\mathcal{R}'_{101}(\omega,i\omega',\omega)$, given by
\begin{eqnarray*}
\mathrm{Re} \{\mathcal{R}'_{101}(\omega,i\omega',\omega)\}  
= {\sum_{\gamma_n}} \frac{(E_n-E_0)
|\langle \delta \phi_{+1} ||D|| \gamma_n,0 \rangle|^2}{(E_n-E_0)^2+(\omega')^2} .
\label{ReR101}
\end{eqnarray*}%
The term $|\gamma_n,0\rangle =|\gamma_0,0\rangle $ automatically disappears because of the $(E_{n}-E_{0})$ in
the numerator.

In practice, the parameter $\omega^{\prime}$ should be chosen sufficiently
small to satisfy the condition $|\omega^{\prime}| \ll |E_0-E_n|$ for any $n$,
but preserve numerical stability when solving the inhomogeneous equation at this $\omega^{\prime}$.
To ensure more flexibility in choosing the parameter $\omega'$, we
orthogonalize $|\delta \phi_q \rangle$ to $|0\rangle $, as
\begin{equation*}
|\delta \phi_q \rangle^{\prime}= \left( 1- |0\rangle \langle 0| \right)\, |\delta \phi_q \rangle,
\end{equation*}
and use $|\delta \phi_q \rangle'$ instead of $|\delta \phi_q\rangle$.

The term $\mathcal{R}_1(q\omega)= (3/2)\,\alpha_{q\omega}$
is calculated following the Eqs.~(\ref{alphaE0})-(\ref{a6}), but
the quantity $\mathcal{R}_1 (\omega,\omega)$ has an additional energy denominator. Therefore, we
 treat this term as the derivative of $\mathcal{R}_1(\omega)$ over $\omega$, i.e.,
\begin{equation*}
\mathcal{R}_1 (\omega,\omega) = \frac{\partial \mathcal{R}_1 (\omega)}{\partial \omega} =
\frac{3}{2}\, \underset{\Delta \rightarrow 0}{\lim} \frac{\alpha_{\omega +\Delta} -\alpha_\omega}{\Delta}.
\end{equation*}
We find $\mathcal{R}_1 (\omega)$ and $\mathcal{R}_1 (\omega,\omega)$
calculating $\alpha_\omega$ and $\alpha_{\omega+\Delta}$, where $\Delta$ is chosen so that $\Delta \ll \omega$.
\paragraph{Method: Sr calculations.}
The electric dipole polarizabilities of the Sr clock states at the magic
wavelength $\lambda^* = 813.4280(5)$ nm~\cite{YeKimKat08} were calculated in our work~\cite{SafPorSaf13}
to be $\alpha_{E1}(\omega^*) = 286.0(3)$ a.u..

In this work we apply the method discussed above to calculate the hyperpolarizabilities and the multipolar polarizabilities, $\alpha^{M1}$ and $\alpha^{E2}$, for the Sr clock states,  $5s^2\, ^1\!S_0$ and $5s5p\, ^3\!P^o_0$ at the magic frequency.
We consider Sr as an atom with two valence electrons above the closed shell core.
The calculations are carried out in the framework of the high-precision
relativistic methods combining configuration interaction (CI) with (i)
many-body perturbation theory ~\cite{DzuFlaKoz96} and (ii)
linearized coupled-cluster  method~\cite{SafKozJoh09}. In
these methods the energies and wave functions are found from the
multiparticle Schr\"odinger equation
\begin{equation}
H_{\mathrm{eff}}(E_n) \Phi_n = E_n \Phi_n,
\label{Heff}
\end{equation}
where the effective Hamiltonian is defined as
\begin{equation}
H_{\mathrm{eff}}(E) = H_{\mathrm{FC}} + \Sigma(E).
\end{equation}
Here, $H_{\mathrm{FC}}$ is the Hamiltonian in the frozen core (Dirac-Hatree-Fock) approximation
and $\Sigma$ is the energy-dependent correction, which takes into account
virtual core excitations in the second order of the perturbation theory (the
CI+MBPT method) or to all orders (the CI+all-order method).

\paragraph{Hyperpolarizabilities.}
The hyperpolarizability is dominated by the valence electrons contribution; the
contribution of the core electrons is small and can be neglected at the present level of accuracy.
In particular, the static hyperpolarizability of the
Sr$^{2+}$ ionic core in the ground state was found in \cite{YuSuoFen15} to be negligible, 62.6 a.u..

To accurately calculate the valence part of the quantities given by Eqs.~(\ref{R_J1}) and (\ref{R_J3}), we
applied solutions of the inhomogeneous equations as was described above.
We use effective (or ``dressed'') operators in our calculations that include the
random-phase approximation (RPA). A concept of effective operators was developed in~\cite{DzuKozPor98}.
\begin{table}[tbp]
\caption{Contributions to the Sr hyperpolarizabilities $\beta(5s^2\,^1\!S_0)$ and $\beta(5s5p\,^3\!P_0^o)$ (in a.u.)
calculated in the CI+all-order (labeled as ``CI+All'') and CI+MBPT (labeled as ``CI+PT'') approximations at the magic frequency.
$\Delta \beta \equiv \beta(^3\!P^o_0) - \beta(^1\!S_0)$ is the difference of the total $^3\!P^o_0$ and $^1\!S_0$
values. Numbers in brackets represent powers of 10. The uncertainties are given in parentheses.}
\label{Tab:beta}%
\begin{ruledtabular}
\begin{tabular}{lcccc}
 & \multicolumn{2}{c}{$5s^2\,^1\!S_0$} & \multicolumn{2}{c}{$5s5p\,^3\!P_0^o$} \\
 Contribution &\multicolumn{1}{c}{CI+All} &\multicolumn{1}{c}{CI+PT} &\multicolumn{1}{c}{CI+All} &\multicolumn{1}{c}{CI+PT} \\
\hline \\ [-0.3pc]
$\frac{1}{9}\mathcal{R}_{101}( \omega, 2\omega, \omega)$        & 5.08[5] & 4.62[5] & -5.96[6] & -5.64[6]  \\[0.1pc]
$\frac{1}{9}\mathcal{R}_{101}($-$\omega,$-$2\omega,$-$\omega)$  & 4.48[4] & 4.38[4] &  9.83[4] &  9.91[4]  \\[0.1pc]
$\frac{1}{9}\mathcal{R}'_{101}( \omega, 0, \omega)$             & 2.41[5] & 2.29[5] &  2.43[6] &  2.68[6]  \\[0.1pc]
$\frac{1}{9}\mathcal{R}'_{101}( \omega, 0,-\omega)$             & 6.47[4] & 6.20[4] &  3.26[5] &  3.47[5]  \\[0.1pc]
$\frac{1}{9}\mathcal{R}'_{101}(-\omega, 0, \omega)$             & 6.47[4] & 6.20[4] &  3.26[5] &  3.47[5]  \\[0.1pc]
$\frac{1}{9}\mathcal{R}'_{101}(-\omega, 0,-\omega)$             & 1.76[4] & 1.70[4] &  6.78[4] &  6.84[4]  \\[0.3pc]

$\mathcal{R}_1( \omega)$                                        & 6.63[2] & 6.49[2] &  6.03[2] &  6.39[2]  \\[0.1pc]
$\mathcal{R}_1(-\omega)$                                        & 1.94[2] & 1.92[2] &  2.67[2] &  2.68[2]  \\[0.1pc]
$\mathcal{R}_1( \omega, \omega)$                                & 1.52[4] & 1.47[4] &  4.79[4] &  5.12[4]  \\[0.1pc]
$\mathcal{R}_1(-\omega,-\omega)$                                & 1.17[3] & 1.17[3] &  2.40[3] &  2.40[3]  \\[0.4pc]

$\frac{1}{9}Y_{101}(\omega)$                                    &-6.18[5] &-6.06[5] & -7.57[6] & -7.50[6]  \\[0.4pc]

$\frac{2}{45}\mathcal{R}_{121}( \omega, 2\omega, \omega)$       & 7.61[5] & 7.10[5] & -1.82[7] & -1.65[7]  \\[0.1pc]
$\frac{2}{45}\mathcal{R}_{121}($-$\omega,$-$2\omega,$-$\omega)$ & 1.83[4] & 1.78[4] &  4.97[4] &  5.03[4]  \\[0.1pc]
$\frac{2}{45}\mathcal{R}_{121}( \omega, 0, \omega)$             & 3.86[5] & 3.68[5] &  9.64[6] &  1.13[7]  \\[0.1pc]
$\frac{2}{45}\mathcal{R}_{121}( \omega, 0,-\omega)$             & 1.07[5] & 1.03[5] &  8.72[5] &  9.76[5]  \\[0.1pc]
$\frac{2}{45}\mathcal{R}_{121}(-\omega, 0, \omega)$             & 1.07[5] & 1.03[5] &  8.72[5] &  9.76[5]  \\[0.1pc]
$\frac{2}{45}\mathcal{R}_{121}(-\omega, 0,-\omega)$             & 2.98[4] & 2.90[4] &  1.44[5] &  1.50[5]  \\[0.4pc]

$\frac{2}{45}Y_{121}(\omega)$                                   & 1.41[6] & 1.33[6] & -6.58[6] & -3.03[6]  \\[0.4pc]

Total                                                           & 7.90[5] & 7.25[5] & -1.42[7] & -1.05[7]  \\[0.2pc]
Recommended                                                     & 7.90(65)[5]  &    & -1.42(37)[7]  &      \\[0.5pc]

$\Delta \beta$                                                  &-1.50[7] &-1.12[7] &          &           \\[0.3pc]
$\Delta \beta$ (Recommended)                                    &\multicolumn{2}{c}{-1.5(4)[7]} & &
\end{tabular}
\end{ruledtabular}
\end{table}

The breakdown of the terms contributing to $^1\!S_0$ and $^3\!P^o_0$ hyperpolarizabilities at the magic frequency
is given in Table~\ref{Tab:beta}. We carry out all calculations using both the CI+MBPT and CI+all-order methods.
The CI+all-order calculations include higher-order terms in comparison with the CI+MBPT calculations and are more
accurate. The difference between these two calculations gives an estimate of the
uncertainty of the results. This method in evaluating uncertainties has been extensively tested for Sr and other
atoms~\cite{SafPorSaf13,PorSafCla14,PorKozSaf16}.

We  compare our recommended value for the differential clock hyperpolarizability at the magic frequency,
$\Delta \beta = \beta(^3\!P_0^o) - \beta(^1\!S_0)$, with the
experimental results of the JILA and SYRTE groups and also with the theoretical value
obtained in~\cite{KatOvsMar15}:
\begin{align*}
\Delta \beta & =-1.5(4)\times 10^{7}\text{ a.u. }(\text{\textrm{this work}}), \\
& =-2.01(45)\times 10^{7}\text{ a.u. }(\text{SYRTE~\cite{WesLodLor11,LeTLorLeC13}}), \\
& =-1.3(1.3)\times 10^{7}\text{ a.u. }(\text{JILA~\cite{NicCamHut15}}), \\
& =-3.74\times 10^{7}\text{ a.u. }(\text{Theory~\cite{KatOvsMar15}}).
\end{align*}
\begin{table}[tbp]
\caption{The dynamic $M1$ and $E2$ polarizabilities (in a.u.) of the $%
5s^2\,^1\!S_0$ and $5s5p\,^3\!P_0^o$ states at the magic frequency, calculated in the
CI+MBPT (labeled as ``CI+PT'') and CI+all-order (labeled as ``CI+All'') approximations.  The recommended
value of $\Delta \alpha^{QM}$ is given in the last line.
The uncertainties are given in parentheses.}
\label{Tab:alpha}%
\begin{ruledtabular}
\begin{tabular}{lccr}
Polariz.                  &    CI+PT              &      CI+All                & \multicolumn{1}{c}{Other theor.}  \\
\hline \\ [-0.5pc]
$\alpha^{M1}(^1\!S_0)$    &  2.19$\times 10^{-9}$ &  2.37$\times 10^{-9}$      &        \\[0.3pc]
$\alpha^{M1}(^3\!P^o_0)$  & -5.09$\times 10^{-6}$ & -5.08$\times 10^{-6}$      &        \\[0.3pc]
$\Delta \alpha^{M1}$      & -5.09$\times 10^{-6}$ & -5.08$\times 10^{-6}$      &         \\[0.5pc]

$\alpha^{E2}(^1\!S_0)$    &  8.61$\times 10^{-5}$ &  8.87(26)$\times 10^{-5}$  &         \\[0.3pc]
$\alpha^{E2}(^3\!P^o_0)$  &  12.1$\times 10^{-5}$ &  12.2(25)$\times 10^{-5}$  &         \\[0.3pc]
$\Delta \alpha_{E2}$      &  3.50$\times 10^{-5}$ &  3.31(36)$\times 10^{-5}$  &          \\[0.5pc]

$\Delta \alpha^{QM}$      &  2.99$\times 10^{-5}$ &  2.80(36)$\times 10^{-5}$  & 0.74$\times 10^{-5}$~\cite{KatOvsMar15} \\
                          &                       &                            &-3.60$\times 10^{-5}$~\cite{OvsPalTai13} \\[0.7pc]
Rec. $\Delta \alpha^{QM}$ &                       &  2.80(36)$\times 10^{-5}$  &
\end{tabular}
\end{ruledtabular}
\end{table}
Our result is  in agreement with the  experimental results, but definitely not consistent with zero.
\paragraph{$M1$ and $E2$ polarizabilities at the magic frequency}
To accurately calculate the valence part of the $E2$ polarizabilities of the clock states
at the magic frequency, given by~\eref{Qlk},  we solved inhomogeneous equation, as described above.
We calculated these quantities using both the CI+all-order and CI+MBPT methods, including the RPA corrections to the quadrupole operator $Q$. The core contributions
were calculated in the RPA approximation. For the $M1$ polarizabilities, only a few low-lying intermediate states give dominant
contributions and it is sufficient to calculate their sum.
We estimate the uncertainties of the results as the difference between the CI+all-order and CI+MBPT values.

The $\alpha^{M1}(^1\!S_0)$ polarizability is very
small and can be neglected. The $\alpha^{M1}(^3\!P^o_0)$  polarizability is
more than 3 orders of magnitude larger, but still an order of magnitude
smaller than $\Delta \alpha^{E2}$. Therefore, the accuracy of
$\Delta \alpha^{QM} \equiv \Delta \alpha^{E2} + \Delta \alpha^{M1}$,
where
\begin{eqnarray}
\Delta \alpha^{M1} &\equiv& \alpha^{M1}(^3\!P^o_0) - \alpha^{M1}(^1\!S_0),  \notag \\
\Delta \alpha^{E2} &\equiv& \alpha^{E2}(^3\!P^o_0) - \alpha^{E2}(^1\!S_0),
\label{delEM}
\end{eqnarray}
is mostly determined by the uncertainty in $\Delta \alpha_{E2}$.

The final values of the polarizabilities, the recommended value of $\Delta \alpha^{QM}$,
and their uncertainties are listed in Table~\ref{Tab:alpha}. We estimate the uncertainty of the recommended value of $\Delta \alpha^{QM}$ to be 13\%.

To conclude, we have developed a method to calculate
hyperpolarizabilities for atoms and ions based on a solution of the inhomogeneous equation, which allows us to carry out complete summations over all intermediate states. We
applied our method to the calculation of the hyperpolarizabilities for the $^1\!S_0$ and $^3\!P_0^o$ clock states in Sr
and found the differential hyperpolarizability to be $-1.5(4)\times 10^{7}$ a.u.
We have also calculated the $M1$ and $E2$ polarizabilities for these states at the magic frequency.

This research was performed under the sponsorship of the U.S. Department of Commerce and National Institute of Standards
and Technology. S.P. and M.K. acknowledge support from Russian Foundation for Basic Research under Grant No. 17-02-00216.


\end{document}